\documentclass[10pt,journal,compsoc]{IEEEtran}

\usepackage[numbers]{natbib}
\usepackage{booktabs}
\usepackage{colortbl}
\usepackage{bm}
\usepackage{graphicx}
\usepackage{subcaption}
\usepackage{cleveref}
\usepackage{soul}

\usepackage{algorithm}
\usepackage{algorithmic}
\usepackage{url}
\usepackage{multirow}
\usepackage{comment}

\usepackage{booktabs}

\usepackage{emerald}
\usepackage{aurical}
\usepackage{pbsi}
\usepackage{calligra}
\usepackage[T1]{fontenc}

\usepackage{tikz}
\tikzstyle{mybox} = [draw=black, very thick, rectangle, rounded corners, inner ysep=5pt, inner xsep=5pt]

\usepackage{array}
\begin{document}

\begin{table*}[]
    \centering
    \large
    \begin{tabular}{>{\centering\arraybackslash}p{\linewidth}} 
    
    \textbf{This work has been submitted to the IEEE for possible publication.}  \\ {Copyright may be transferred without notice, after which this version may no longer be accessible.}  \\ 
   
       \end{tabular}

\end{table*}

\newpage

\title{The Role of Cognitive Abilities in Requirements Inspection: Comparing UML and Textual Representations
\thanks{Research supported by Grant PID2022-137846NB-I00, funded by MCIN/AEI/10.13039/501100011033 and by ERDF A way of making Europe. The authors would like to thank all the participants of the study.}}                     



\author{Giovanna Broccia, Sira Vegas, and Alessio Ferrari%
\IEEEcompsocitemizethanks{%
\IEEEcompsocthanksitem G.~Broccia is at 
CNR--ISTI, Pisa, Italy. 
\IEEEcompsocthanksitem S.~Vegas is at Universidad Politecnica de Madrid, Spain. 
\IEEEcompsocthanksitem A.~Ferrari is at University College Dublin, Ireland. 
\IEEEcompsocthanksitem Corresponding Authors: G.~Broccia and A.~Ferrari.\\ E-mail: \
giovanna.broccia@isti.cnr.it,
alessio.ferrari@ucd.ie}}

\markboth{Submitted to IEEE Transactions on Software Engineering, 2026}%
{Broccia \MakeLowercase{\textit{et al.}}: The Role of Cognitive Abilities in Requirements Inspection: Comparing UML and Textual Representations}






\IEEEtitleabstractindextext{%
\begin{abstract}
\textit{Context.}  
The representation of requirements plays a critical role in the accuracy of requirements inspection. While visual representations, such as UML diagrams, are widely used 
alongside text-based requirements, 
their effectiveness in supporting inspection is still debated.
Cognitive abilities, such as working memory and mental rotation skills, may also influence inspection accuracy. 
%
\textit{Objective. }
This study aims to evaluate whether the use of UML sequence diagrams alongside text-based requirements improves the accuracy of requirements inspection compared to text-based requirements alone and to explore whether cognitive abilities are associated with differences in performance across the two treatments (text \textit{vs} text with UML support).
\textit{Methods.}
We conducted a crossover experiment with 38 participants
to assess the accuracy of requirements inspection under the two treatments in terms of issues found and justifications provided. 
Linear mixed-effects and generalized linear models were used  to analyse the effects of 
treatment, period, sequence, and cognitive abilities.
\textit{Results. }
The results indicate a significant three-way interaction between representation type, working memory capacity, and mental rotation ability. This finding suggests that the effectiveness of UML support is not uniform across individuals: participants with high scores in both cognitive abilities experienced reduced performance when using UML for violation detection. Conversely, the same cognitive profile was associated with improved justification accuracy under UML-aided inspection, indicating that higher cognitive abilities may support deeper reasoning processes when dealing with multi-modal information, i.e., diagrams and text. 
%
%
%
\textit{Conclusion. }
The findings suggest that while UML diagrams may not universally improve requirements inspection accuracy, 
certain cognitive profiles can benefit from them to improve inspection performance. 
These results highlight the importance of considering individual cognitive profiles when designing and assigning requirements inspection tasks, and that multiple means of representation are not always beneficial. 

\end{abstract} }


\maketitle

\section{Introduction}
Software inspection refers to the systematic review of software artifacts by trained professionals using a well-defined process to identify and address defects This method can uncover and correct between 50\% and 90\% of defects at any stage of development \cite{Fagan1986AdvancesIS,10.5555/562741}, and reduce maintenance costs by a factor of 10 to 100 \cite{5388086}. 
Various inspection processes exist, with Fagan’s method being one of the most established and widely used \cite{Fagan1986AdvancesIS}. Within these processes, inspectors can adopt different \textit{reading techniques}~\cite{zhu2016software}. These are systematic strategies that guide how software artifacts are reviewed, offering different levels of structure and focus during violations identification. Examples of techniques include ad-hoc reading, checklist-based reading (CBR), or perspective-based reading (PBR).

Inspection \textit{performance} can be influenced by multiple factors. Among these are (i) the reading technique used, (ii) the representation of the software artifact, and (iii) the inspectors' abilities and experience. Prior empirical research has examined these factors, but predominantly in isolation.

For reading techniques (i), previous  research demonstrates that the selected method can affect both the number and nature of defects detected and the overall inspection efficiency 
\cite{laitenberger2000experimental,sabaliauskaite2002experimental,rong2012effect,porter1995comparing}.
Regarding software artefact representation (ii), 
prior studies indicate that graphical representations, such as UML diagrams, can enhance defects detection during requirements inspection 
\cite{albayrak2009experiment,ghafory2021experimental}. 
Inspector abilities (iii), also play a crucial role in determining inspection outcomes. Skilled inspectors tend to be more proficient at detecting complex defects, assessing their severity, and conducting inspections efficiently \cite{biffl2000analysis,carver2008impact}. Beyond the inspection domain, a substantial body of research in cognitive psychology suggests that, besides technical skills, also \textit{cognitive abilities} contribute to performance in complex analytical tasks. In particular, working memory (WM) capacity has been shown to influence task success in several domains~\cite{daneman1980individual,miyake1994working,barrouillet1996transitive}, while spatial ability has been associated with improved performance across a wide range of problem-solving tasks~\cite{raghubar2010working,sharafi2021toward,mansoor2023empirical}.
Despite this growing evidence, the explicit role of such cognitive abilities in requirements inspection remains under-explored.
Among the few empirical studies investigating this topic, Sharif et al. \cite{sharifexamining} analysed the effects of inspectors’ cognitive abilities and UML class diagram layouts on defects detection performance. 
The authors considered  different abilities, i.e., WM 
capacity and mental rotation, and different UML representations as isolated factors.   
However, these aspects do not operate independently. Instead, they interact in ways that can jointly shape inspection performance. For instance, graphical representations may either support or hinder cognitive processing depending on how well they align with users’ cognitive abilities. Well-designed graphics can compensate for limitations such as restricted WM capacity \cite{tory2004human}, while poorly designed ones can impose additional cognitive load and reduce effectiveness \cite{huang2009measuring}. 





In this paper, we present a crossover experiment aiming at investigating how representation format and cognitive abilities \textit{jointly} shape inspection performance. Specifically, we examine whether the use of UML sequence diagrams improves the accuracy of requirements inspection, and how this effect depends on participants' WM capacity and mental rotation ability. In our experiment, we fix the reading technique to CBR, because, compared to other techniques, it provides a uniform procedure improving treatment fidelity and reducing variance caused by differing personal strategies. 
%
%
To evaluate inspection performance, we focus on both the ability to identify requirements that violates a quality checklist item and the accuracy of justifying the violation. 
We consider WM capacity and mental rotation ability because these are theoretically linked to key aspects of requirements inspection: holding and integrating information (WM capacity) and interpreting visual structures such as UML diagrams (spatial ability). These abilities are also commonly examined in software engineering research.

%
Our findings reveal that the combined influence of UML use, WM capacity, and mental rotation ability significantly shapes inspection performance: when these factors are analysed jointly, they meaningfully moderate how UML affects both violations detection and justification accuracy. In particular, their interaction decreases violation detection accuracy, but improves justification accuracy.
In contrast, when UML support, WM capacity, and mental rotation are analysed separately, some effects are not statistically significant, and in some cases the observed trends move in the opposite direction. This indicates that conclusions drawn from isolated factors can be misleading, as they fail to capture how these abilities jointly modulate UML’s impact on inspection performance.
These findings underscore the need to consider individual cognitive profiles when introducing visual notations like UML into inspection tasks.

The main contributions of this work are:
\begin{itemize}
    \item A novel empirical investigation of the interaction between cognitive abilities and requirements representation format in the context of requirements inspection.
    \item Evidence that cognitive abilities moderate inspection performance differently depending on the outcome performance measure (violation identification vs. justification accuracy).
    \item Practical insights for tailoring inspection techniques and tool support based on participants’ cognitive profiles, particularly in educational or training settings.
\end{itemize}


The remainder of the paper is structured as follows. Section \ref{sect:backgroundRelated} provides the background on cognitive abilities and requirements inspection, and addresses the related work. Section \ref{sect:studyDesign} presents the experimental design. The results and discussion are presented in Sections \ref{sect:results}, \ref{sect:postHoc}, and Section \ref{sect:discussion}, respectively. Section \ref{sect:threats}  addresses the threats to validity. Finally, Section~\ref{sect:conclusion} summarises the conclusions and suggests directions for future research.

\smallskip
\noindent
\textbf{Replication Package.} We made our replication package available at \cite{broccia_2025_17649266}.

\section{Background and Related Work} \label{sect:backgroundRelated}

\subsection{Requirements Inspection}
Software inspection encompasses a family of structured review processes aimed at detecting quality aspects violations in software artefacts. Among the various approaches proposed in the literature, Fagan’s inspection method remains one of the most influential and widely adopted \cite{Fagan1986AdvancesIS}.
This method typically includes a planning phase, a presentation phase---where the author introduces the artifact---and an individual inspection phase---during which inspectors review the artifact independently to detect violations. The process concludes with an inspection meeting where inspectors discuss their findings, often to resolve false positives rather than uncover new issues.
The individual inspection phase can be carried out using 
different \textit{reading techniques}~\cite{zhu2016software}. 
They are techniques that guide the inspection by providing structured or unstructured methods for reviewing software artifacts. 
These artifacts can include requirements specifications, design documents, or source code.

Reading techniques vary in their level of structure and guidance. In \textit{ad-hoc reading} the detection process is unspecified and driven largely by the inspector's interests and experience; \textit{checklist-based reading} (CBR), entails each inspector focusing on a list of quality aspects following a checklist as a comprehensive guide, ensuring that key criteria---such as coding standards, design principles, and common defect patterns---are thoroughly considered during the review process;  in \textit{perspective-based reading} (PBR), the inspectors assume the viewpoint of different users or stakeholders---such as software designers, testers, or end-users---based on specific scenarios or perspectives to ensure coverage 
(see  \cite{zhu2016software} for details on reading techniques).

\subsection{Cognitive Backgound}\label{sect:background}
\textbf{Working Memory. }
Working Memory is a cognitive system with limited capacity, often referred to as \emph{WM span} or \textit{capacity}, that is responsible for temporarily holding and processing information essential for task completion~\cite{miller1956magical}. WM serves to retain the necessary information for a brief period while concurrently processing other relevant data required to successfully execute the task.

WM capacity was proven to be highly predictive of performance in processing ambiguous syntactic constructions \cite{miyake1994working}, reasoning~\cite{barrouillet1996transitive}, and programming skills~\cite{mansoor2023empirical}, and a key factor in overall intellectual ability~\cite{conway2002latent, conway2003working}.

To measure WM capacity, specific tasks known as WM span tasks (WMSTs) are used~\cite{conway2005working}. These tasks are designed to measure WM performance, requiring the maintenance and recall of specific information, typically numbers, while concurrently performing a processing activity, e.g., evaluating the correctness of equations.

\medskip\noindent
\textbf{Rotation Ability. }
Rotation Ability is a specific aspect of spatial ability that involves the capacity to visualize and manipulate objects in a three-dimensional space. This cognitive skill is crucial for tasks that require the mental transformation of objects, such as understanding diagrams, solving puzzles, and navigating environments.
Research suggests that spatial ability is associated with mathematical problem-solving skills \cite{hegarty1999types, raghubar2010working}, data structure manipulation \cite{sharafi2021toward}, and the understandability of formal notations \cite{mansoor2023empirical}.

To assess rotation ability, tasks such as the mental rotation task are commonly employed. In this task, participants are presented with pairs of three-dimensional objects and asked to determine whether they are identical or mirror images of each other. 


\subsection{Related Work}\label{sec:relatedWork}
\subsubsection{Empirical Comparisons of Reading Techniques}\label{sect:RelWorComparis}

Numerous studies have compared inspection techniques for violations detection, especially for UML documents. In this section, we focus on studies that employed CBR, which is the technique adopted in our experiment. We do not attempt to provide a comprehensive review of all reading techniques.

Laitenberger et al. \cite{laitenberger2000experimental} conducted an experiment with 
practitioners comparing PBR 
to CBR, for defect detection in UML designs. 
The results show that PBR is more effective and cost-efficient than CBR for defect detection in UML design documents. 
A similar study was conducted by
Sabaliauskaiten et al. \cite{sabaliauskaite2002experimental}  that conducted a controlled experiment with 
students. Their results show that defect detection effectiveness is comparable between the two techniques (PBR: 69\%, CBR: 70\%). However, although reviewers using PBR complete the inspection in less time, the cost per defect is lower for reviewers using CBR.

Mendonça et al. \cite{mendoncca2008framework} describes the replication of a number of experiments, including those originally conducted by Laitenberger et al. \cite{laitenberger2000experimental} 
The results of these replications generally supported the effectiveness of PBR over CBR, showing that PBR was more effective and efficient in detecting defects, although some variations were observed depending on the specific context and subject experience. 

Rong et al. \cite{rong2012effect} examined the effectiveness of ad-hoc reading vs. CBR in code review 
conducting a semi-controlled experiment with
first-year software engineering students.
The results indicated that while CBR helped guide the review process and reduced the review rate, they did not significantly improve the review efficiency in terms of defect detection. 
Porter et al. \cite{porter1995comparing} conducted an experiment with graduate students to compare the effectiveness of different reading techniques for software requirements inspections. 
The results showed that the scenario-based method had a higher fault detection rate compared to ad hoc and CBR methods. 

These studies illustrate that structured reading techniques like PBR might offer advantages in structured inspections, especially in complex documents, while the structured nature of CBR could make it more suitable for non-expert inspectors. 
We used CBR in our experiments as it is more suitable for non-expert users.

\subsubsection{Empirical studies on UML in requirement inspection
}\label{sect:relWor_UML}

The role of UML sequence diagrams in enhancing requirements inspection has been explored in several studies. Albayrak \cite{albayrak2009experiment} conducted an experiment to  examine the impact of UML diagrams on defect reporting. Results indicated that UML inclusion increased the total number of detected defects but did not enhance the number of correctly detected defects.
%
Ghafory \cite{ghafory2021experimental} conducted an experiment 
to evaluate the effectiveness of integrating use case and activity diagrams in requirements specifications on the 
time taken to inspect the documents 
and the number of reported faults. 
The results indicated that students inspecting documents with text only reported more faults overall, but also more incorrect faults, compared to those inspecting documents with text and diagrams. The number of correct faults reported was similar between both groups, but those inspecting documents with diagrams did so in less time.

 These studies collectively highlight the potential of UML diagrams to enhance the defect detection process in requirements inspections. Our study builds on this by specifically focusing on sequence diagrams and their impact on inspection accuracy, providing a more targeted investigation into the role of this particular type of UML diagram. Additionally, we examine the interaction between the use of UML diagrams and inspectors’ cognitive abilities to understand how these factors together influence inspection outcomes.

\subsubsection{Empirical studies on cognitive factors in inspection performance} \label{sec:RelWorAbilities}
Although cognitive abilities have been shown to influence performance in several software engineering tasks, their role in requirements inspection has received very limited attention. 
Sharif et al. \cite{sharifexamining} conducted a controlled experiment to investigate the effects of UML layouts (multi-cluster vs. orthogonal) and inspectors' cognitive abilities (WM capacity and mental rotation) 
on defect identification. Unlike our study, which explicitly models the joint interaction between cognitive abilities and representation format, Sharif et al. examined each cognitive measure independently and did not assess combined or higher-order effects. 
The results indicated that cognitive abilities were weakly and inconsistently correlated with the performance of the defect detection tasks, and
no significant differences were found between layouts w.r.t. defect identification accuracy and time.

Our study departs from this line of work by explicitly modelling how two cognitive abilities---WM capacity and mental rotation ability---act in combination when inspectors analyse UML-supported requirements.
Additionally, by examining not only whether inspectors detect requirement violations, but also how accurately they justify them, we reflect a more realistic inspection scenario, e.g., in safety-critical domains~\cite{ferrari2018detecting},  in which the identification of a violation is typically associated with an explanation of the rationale, so that the requirements editor can get actionable recommendations to solve the violation.

\section{Study Design}\label{sect:studyDesign}

\subsection{Goal and Research Questions}

Our experimental design follows the guidelines by Vegas et al. \cite{vegas2015crossover} to conduct crossover experiments. The overall goal of the experiment is as follows:
\smallskip

\hspace*{-5mm}
\begin{tikzpicture}
\node [mybox] (box){%
    \begin{minipage}{.95\columnwidth}
    \centering
To investigate how \textbf{different cognitive abilities} (mental rotation and WM capacity) and the \textbf{use of specific requirements representations} (UML vs. textual descriptions) \textbf{influence 
requirements inspection performance} both in terms of detecting quality issues and justifying them, while also examining the interactions between these factors.

    \end{minipage}
};
\end{tikzpicture}%

To achieve the goal our experiment aims to answer the following research questions (RQs):

\begin{description}
    \item[\textbf{RQ1}] What is the influence of treatment (UML+textual requirements vs. textual requirements) on issue detection accuracy, across different levels of mental rotation ability and WM capacity?
    \item[\textbf{RQ2}] What is the influence of treatment (UML+textual requirements vs. textual requirements) on issue justification accuracy, across different levels of mental rotation ability and WM capacity?

\end{description}

\subsection{Variables}
\subsubsection{Independent Variables}
The experiment manipulates one independent variable: the representation format of the requirements. Two treatments are used:
\begin{itemize}
    \item \textbf{Textual requirements}: Participants inspect requirements presented exclusively in textual form.
    \item \textbf{UML-supported requirements}: Participants inspect requirements presented in textual form accompanied by a UML sequence diagram that represents the system described in the text.
\end{itemize}

\subsubsection{Dependent Variables} 
The inspection outcome measures in this study are organised into two overarching constructs, each capturing a distinct aspect of performance during requirements inspection. The first construct, \textit{issue detection accuracy}, reflects participants’ ability to correctly identify quality checklist item violations (which we refer to as \textit{issues}) and is measured using \textbf{\textsl{F1-score}} (a standard information retrieval metric)\footnote{While the more accurate $F_{\beta}$ would be more appropriate, the computation of the $\beta$ value requires further analysis, which we leave for future work.}. \textsl{F1-score} measures the harmonic mean of two complementary measures, namely precision and recall: precision measures the fraction of correctly identified violations among all identified violations; recall measures the fraction of correctly identified violations among all actual violations.

The second construct, \textit{issue justification accuracy}, captures the participants’ ability to clearly and correctly explain why a given requirement violates a checklist item. This is measured through a custom variable, \textbf{\textsl{Accuracy Why}}, which assesses the quality of the justification provided for each correctly identified violation.

\subsubsection{Covariates}\label{sect:covariates}
Two covariates are considered to assess their influence on the dependent variables: \textit{mental rotation ability}, represented by a variable we refer to as \textsl{\textbf{rotation score}}, and \textit{working memory capacity}, denoted as \textsl{\textbf{WM capacity}}. 

\begin{figure*}[t]
    \centering

    \begin{subfigure}[t]{0.3\linewidth}
        \centering
        \includegraphics[width=\linewidth]{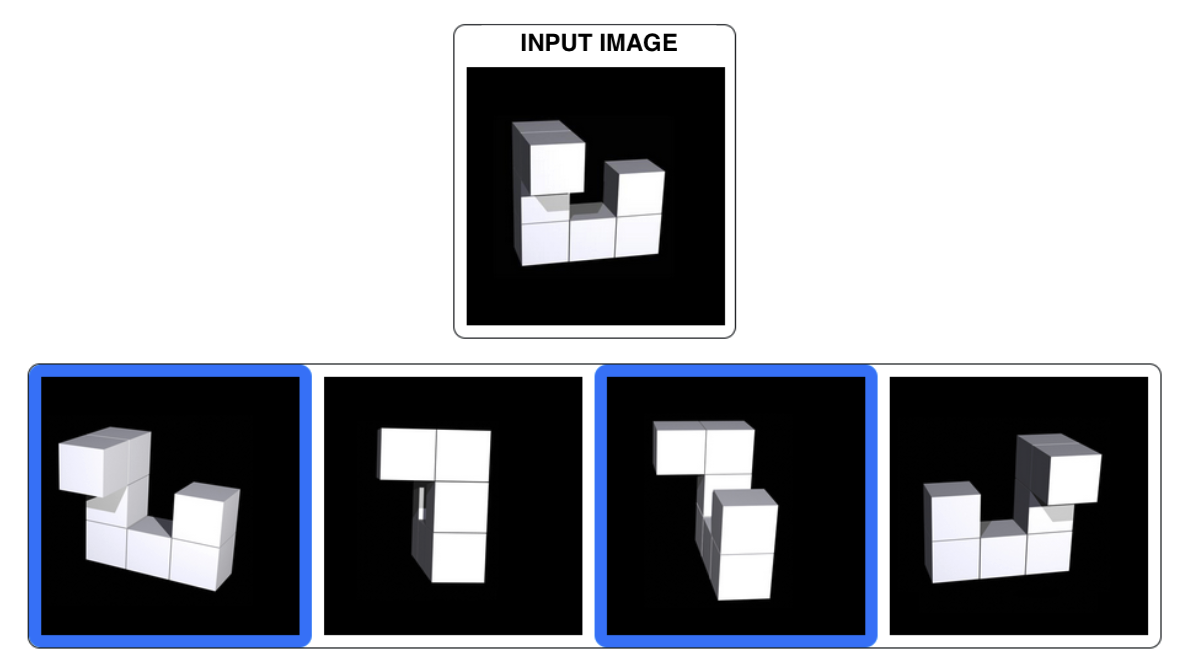}
        \caption{3D Mental Rotation Task.}
        \label{fig:mentalRotation}
    \end{subfigure}
    \hfill
    \begin{subfigure}[t]{0.65\linewidth}
        \centering
        \includegraphics[width=\linewidth]{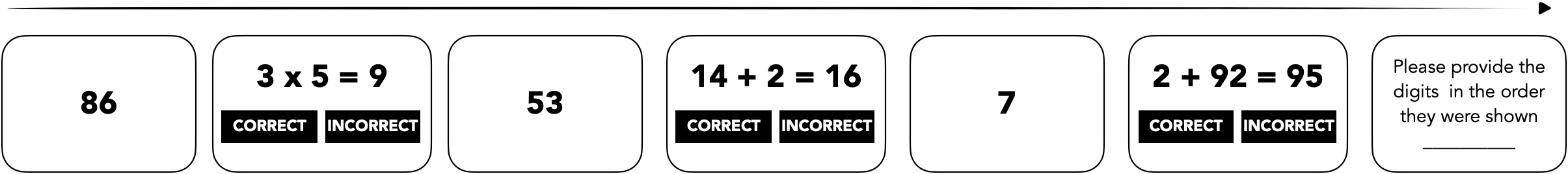}
        \caption{Operation Span Task.}
        \label{fig:ospan}
    \end{subfigure}
        \caption{Cognitive Tasks.}
\end{figure*}

These covariates were derived from two standard cognitive tasks administered prior to the inspection activity. 
3D mental rotation task~\cite{vandenberg1978mental} is a cognitive task designed to evaluate participants' capacity to recognise the spatial rotation of 3D objects. 
For its completion, participants are required to complete 20 trials. In each trial, they are presented with a set of four images depicting 3D objects and must identify which of these images represents a correct rotation of a given input image. Each participant must recognise two correct rotations for each input image.
An example of a trial is illustrated in Figure
\ref{fig:mentalRotation}.
Operation span task~\cite{turner1989working} is a cognitive task commonly used to assess the capacity of users' WM.
To complete it, participants must undergo 12 trials, each varying in complexity and length. In each trial, participants are presented with a set of 2, 3, 4, or 5 numbers, with each set length appearing three times across the total trials (i.e., 4 set length $\times$ 3 repetitions = 12 trials). The numbers are displayed on the screen for one second, interleaved with equations that participants need to evaluate as either correct or incorrect. While assessing the correctness of each equation, participants must memorise the numbers. Once all numbers and their corresponding equations have been presented, participants are prompted to recall the numbers in the order they were shown.  Figure \ref{fig:ospan} illustrates an exemplified sequence of actions required in a trial involving three numbers.

\subsection{Selection of Subjects}
Participants in this study were opportunistically recruited based on their availability, and participation was entirely voluntary. They were informed that the collected data would be used for research purposes in an anonymous and aggregated form, and that no link between participant identity and responses was retained.
They were master’s students in Software Engineering at the Universidad Politécnica de Madrid (UPM), enrolled in the Verification and Validation course. As part of the course, students received training in  reading techniques. 

\subsection{Choice of Design}
\begin{figure}
    \centering
    \includegraphics[width=1\linewidth]{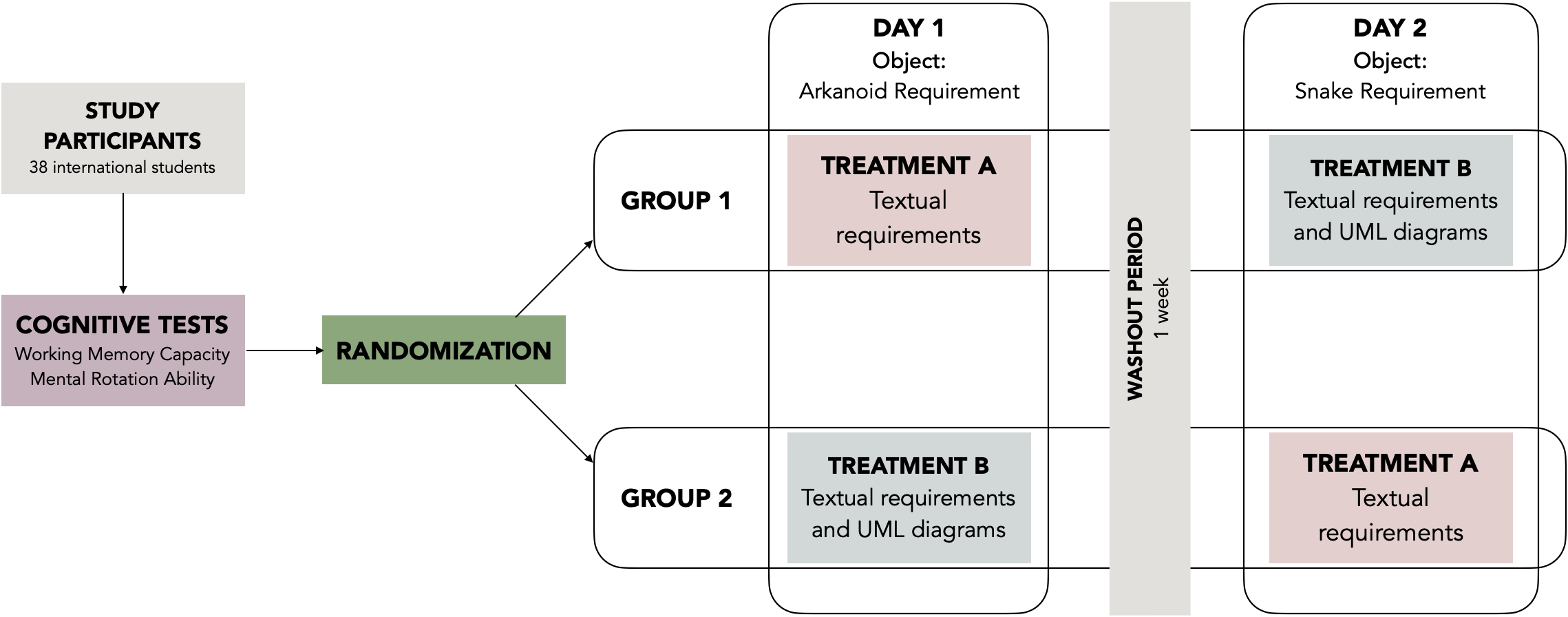}
    \caption{Crossover Experimental Design. 
    }
    \label{fig:studyDesign}
\end{figure}

To achieve this study's objectives, we conducted a crossover experiment with 
participants divided into two groups (\textit{Group 1} and \textit{Group 2}), balanced in terms of cognitive abilities (i.e., WM capacity and mental rotation ability), which were assessed in advance through two cognitive tasks (cfr. Section \ref{sect:covariates}). Participants were assigned to groups using stratified randomisation, based on clusters derived from their cognitive ability levels, to ensure comparability between groups. Each group performed requirements inspections under two treatments and sequences: Group 1 began with text-based requirements (\textit{Treatment A}) and then proceeded to UML-supported requirements (\textit{Treatment B}), while Group 2  followed the reverse sequence. The inspections took place over two days (Day 1 and Day 2) with a one-week washout period in between, ensuring minimal carryover effects. Each day featured a distinct requirements document, describing the implementation of a video game---Arkanoid on \textit{Day 1} and Snake on \textit{Day 2}. 
Figure \ref{fig:studyDesign} shows such a crossover design.

\subsection{Experimental Procedure}\label{sect:phases}

The study consists of four phases
:
\begin{enumerate} 
\item \textbf{Introductory phase. } Participants began with a 2-hour introductory lesson that 
covers topics such as requirement inspection, checklist-based evaluation, cognitive aspects, and an overview of the entire experimental process and materials. 
They also engaged in a guided inspection task using a sample requirement document, which is similar yet shorter than the ones used during the experimental phase (cfr. Section \ref{sec:object})
, ensuring that participants are well-prepared for the inspection tasks ahead. 

\item \textbf{Cognitive Phase. } 
Participants were asked to complete the cognitive tasks to assess mental score and WM capacity. 
They had the flexibility to perform these tasks at their convenience, ideally when they felt most cognitively alert. 
Participants were instructed to complete both cognitive tasks no later than three days before the next phase of the study---to enable us to cluster the subjects considering their cognitive abilities. 

\item \textbf{Inspection phase. } 
Participants were asked to complete an inspection task over two sessions, on Day 1 and Day 2, under two different treatments: one with text-based requirements and another with UML-supported requirements, with a one-week washout period.

The inspection task entails analysing system requirements containing a number of pre-existing, injected violations. Regardless of the treatment used, participants were instructed to refer to a provided requirements quality checklist and to fill a test document (cfr. Section \ref{sec:object}). They were instructed to identify any requirement that violates a checklist item and to provide a justification for each identified violation.
Under Treatment B (UML-supported inspection), they were asked to indicate whether they relied on the textual description, the UML diagram, or both when identifying each violation.


\item \textbf{Post-test questionnaire.} 
Participants completed an online questionnaire assessing their 
experience with software inspection, their preference between the two treatments, their background in technical drawing, and any prior experience with tasks similar to the mental rotation test.
The data collected in this phase provided valuable feedback, helping to confirm or challenge our findings from the inspection phase.

\end{enumerate}

\subsection{Experimental Objects}\label{sec:object}
During the inspection tasks
, participants 
received the requirement documents to be inspected, a UML sequence diagram of the system under analysis (when performing the task under Treatment B), a checklist document to guide the inspection process, and a test document to complete the task.

The \textbf{requirements documents} used during the two sessions of the inspection phase (i.e., Day 1 and Day 2) describe two distinct video game implementations: 
the Arkanoid game\footnote{\url{https://en.wikipedia.org/wiki/Arkanoid}} on Day 1; and the Snake game\footnote{\url{https://en.wikipedia.org/wiki/Snake_(1998_video_game)}} on Day 2. 
Each document includes a short system description (not subject to inspection) and a set of requirements that were deliberately injected with a set of checklist items violations  (Table \ref{tab:reqs} provides details on the number of requirements and issues in each document). 
A single requirement may generate multiple issues. For instance, the following requirement:

\textit{The system shall calculate the optimal route and display it quickly using the available data.}

generates two issues, each corresponding to a violated checklist item:
Issue 1 -- Violation of checklist item 2 (clarity and unambiguity):
The terms ``optimal'' and ``quickly'' are subjective and undefined, making the requirement ambiguous; Issue 2 -- Violation of checklist item 6 (sufficient inputs):
The requirement does not specify which input data are needed to compute the route, preventing correct interpretation and implementation.


\begin{table}[t]
    \centering
    \begin{tabular}{c|c|c}
         & \textbf{\#requirements} & \textbf{\#issues}\\
      Arkanoid Game  & 10 & 96\\
       Snake Game  & 9 & 87\\
    \end{tabular}
    \caption{Number of requirements and issues for each requirement document object.}
    \label{tab:reqs}
\end{table}

The \textbf{quality checklist} used during the inspection task is shown in Table \ref{tab:checklist} and it is a simplified version of the checklist used in \cite{porter1995comparing}.

The \textbf{test document}, an excerpt of which is shown in Table \ref{tab:testDoc},
lists each checklist item in a separate row.
For every checklist item, participants were instructed to report each issue by:
(i) identifying the requirement(s) that violate the item (and mark it in the WHERE column); (ii) providing a brief justification for the violation (and mark it in the WHY column).
%
The WHERE entry could involve a single requirement, all requirements indicating that the violations affects the entire requirements document, or it could pertain to a pair of two specific requirements.

The \textbf{post-test questionnaire} was conducted online. Participants received the link to the questionnaire via email.


\begin{table}[]
    \centering
    \begin{tabular}{c p{0.85\columnwidth} }
    \hline
    \textbf{ID} & \textbf{ITEM} \\ \hline
    1 & All the goal of the system are defined. \\ \hline
    2 & The requirements are clear and unambiguous. \\ \hline
    3 & Each functional requirements specify input and output, as well as function, as appropriate. \\ \hline
    4 & The requirements provide an adequate basis for system design. \\ \hline
    5 & The described functions are sufficient to meet the system objectives. \\ \hline
    6 & All the inputs to a function are sufficient to perform the required function. \\ \hline
    7 & The undesired events are considered and their required responses are specified. \\ \hline

    8 & The individual requirements are stated so that they are discrete, unambiguous, and testable. \\ \hline
    9 & All transitions are specified deterministically. \\ \hline
    10 & The requirements are mutually consistent. \\ \hline
    11 & The requirements are free of duplication and conflict with other requirements. \\ \hline
    12 & Each requirement has only one interpretation. If a term could have multiple meaning, it is defined. \\ \hline
    13 & All the described functions are necessary to meet the system objectives. \\ \hline
    14 & All inputs to a function are necessary to perform the required function. \\ \hline
    15 & All the outputs produced by a function is used by another function or transferred across objects/subjects. \\ \hline

       \end{tabular}
    \caption{Checklist document used for the inspection task. }
    \label{tab:checklist}
\end{table}

\begin{table*}[]
    \centering
    \begin{tabular}{|p{0.35\linewidth} | p{0.15\linewidth} |p{0.4\linewidth} |}
    \hline
    \textbf{ITEM} & \textbf{WHERE} {\scriptsize For items that are not satisfied specify where (e.g. ALL, R1, R2). }& \textbf{WHY} {\scriptsize For items that are not satisfied specify why.}  \\ \hline 
  All the goal of the system are defined. & \texttt{All} & \texttt{The goal of the system is not specified}\\ \hline
    \multirow{3}{*}{The requirements are clear and unambiguous.} & \texttt{R1}  & \texttt{The shape of the game board is not clearly specified} \\ 
    & \texttt{R1}  & \texttt{The size of the game board is not clearly specified}\\ 
    & $\dots$ & $\dots$\\ \hline
  
       \end{tabular}
    \caption{Excerpt of the inspection test document with example answers.  
    }
    \label{tab:testDoc}
\end{table*}

\subsection{Data Collection and Measurement Procedure}

\subsubsection{Inspection Task}
Based on the known issues, we first established a preliminary ground truth. 
However, this initial ground truth was not exhaustive. After reviewing participants’ inspection responses, we refined and expanded it by incorporating additional valid issues they correctly identified but were not previously included\footnote{Although the issues were injected by requirements experts, this does not guarantee that all issues were captured, as the collective effort of multiple inspectors typically results in identifying a greater number of violations \cite{aurum2002state}, and our goal was to construct a comprehensive ground-truth.}. The resulting ground truth is a test document in which the “WHERE’’ column lists all requirements that violate a given checklist item, and the “WHY’’ column provides the corresponding justification.

\textsl{F1-score} is computed using the count of true positives (TPs), false positives (FPs), and false negatives (FNs). 
A TP occurs when the participant correctly identifies an issue (i.e., correctly marks the WHERE) \emph{and} provides a justification (WHY) that is at least partially relevant.
FPs arise either when a participant indicates a requirement that does not violate the checklist item, or when the WHY justification is unrelated to the true reason for the violation.
FNs correspond to issues present in the ground truth that the participant fails to report.

\textsl{Accuracy Why} evaluates the quality of the participant’s justifications (the WHY column).
Two authors independently scored each WHY using a 0–4 Likert scale that measures the relevance and clarity of the motivation (cfr. Table~\ref{tab:likert}), resolving disagreements by discussion.

The identification and count of TPs, FPs, and FNs are performed as follows: for each checklist item, every WHERE provided by the participant was matched against the ground truth.

\begin{itemize}
    \item If the WHERE correctly corresponds to a true issue, the associated WHY is scored on the Likert scale. A score $>0$ represents a motivation that, in different measure, is relevant to the real motivation for which the requirement is violating the checklist item, and categorises the response as a TP; a score of $0$ (unrelated motivation)  categorises it as a FP.
    \item If the WHERE does not correspond to any true issue for that checklist item, it is marked as a FP.
    \item Any true issues listed in the ground truth that were not identified by the participant are counted as FNs.
\end{itemize}

\begin{table}[]
    \centering
    \begin{tabular}{p{0.07\columnwidth} | p{0.2\columnwidth} | p{0.5\columnwidth}}
        \textbf{Score}  & \textbf{Value} & \textbf{Explanation} \\ \hline
        0 & \textbf{Not a correct motivation} & The motivation is entirely unrelated to the error in the requirement.\\ \hline
        1 &  \textbf{Motivation somewhat unclear} & The motivation is weakly presented, with partial relevance to the error but lacks clarity or depth.  \\ \hline
        2 & \textbf{Motivation somewhat clear} & The motivation is mostly relevant and understandable, but some details are missing or underdeveloped. \\ \hline
        3 &  \textbf{Motivation clear} & The motivation is relevant and well-presented, though with minor details missing or some room for improvement. \\ \hline
        4 & \textbf{Motivation clearly presented} & The motivation is fully clear, precise, and strongly aligns with the error in the requirement, leaving no ambiguity.\\ \hline
    \end{tabular}
    \caption{Likert scores explanation.}
    \label{tab:likert}
\end{table}

\medskip
\textsl{Accuracy Why} is computed as the median of the Likert scores assigned to all TPs for each participant.

\subsubsection{Cognitive Tasks}
The cognitive phase was conducted online across two different web platforms: one for the WM capacity assessment\footnote{\url{https://wmspantasks.isti.cnr.it/operationSpan.html}} and one for the mental rotation ability assessment\footnote{\url{https://3dmentalrotation.isti.cnr.it}}.
The latter is a web-based refactoring of the Java app used in the study presented in \cite{mansoor2023empirical}. 
Both applications save the results from the cognitive tasks in a CSV file.

\noindent \textbf{Rotation score.} 
The application is developed in order to randomly select two correct rotation images and two incorrect rotation images for each of the 20 input images. Resulting in 40 rotation images to be correctly identified.

\textsl{Rotation score} is computed as the ratio of correctly identified images to the total number of correct images available.
This computation yields a score ranging from 0 to 1.


\noindent \textbf{WM capacity.} 
The evaluation of WM span tasks is generally performed only for the mnemonic activity (i.e., how much the user was good in memorising the numbers)~\cite{kane2004generality,waters1996measurement}, regarding the processing activity (i.e., evaluating the correctness of the equations) as a distracting factor. Different scoring methods exist, e.g. all-or-nothing scoring, partial-credit scoring, and edit-distance scoring \cite{gonthier2023easy}. In our study, we use the partial-credit unit (PCU) scoring method (i.e. the mean proportion of digits within a trial that were recalled correctly), which is favoured by empirical results~\cite{conway2005working}.

The PCU for each user is computed as follows:

\begin{equation}
    \textit{PCU} = \frac{\sum_{i=1}^N \frac{b_i}{a_i}}{N}
\end{equation}

\noindent where $N$ is the number of trials, $b_i$ the number of elements correctly recalled in the trial, and $a_i$ the number of elements to recall in the trial \cite{navarro2024item}. 
PCU scores range from 0 to 1.

\begin{table*}[hbt!]
\renewcommand{\arraystretch}{1.2}
\caption{Hypotheses for each dependent variable. \\
\scriptsize The index $i$ denotes the dependent variable under analysis, namely \textsl{F1-score} (FS) or \textsl{Accuracy Why} (AW). For each variable, H$_{i0}$ represents the null hypothesis, while H$_{i1}$ to H$_{i8}$ represent the corresponding alternative hypotheses. Hypotheses are instantiated using the prefix of the dependent variable, for example H$_{FS1}$ for \textsl{F1-score} and H$_{AW1}$ for \textsl{Accuracy Why}.}
\centering
\scriptsize
\setlength{\tabcolsep}{4pt}
\begin{tabular}{|c|p{15cm}|}
\hline

\textbf{H}$\bm{_{i0}}$ & There is no difference in the effect of treatment (UML-supported requirements vs. textual requirements) across different levels of mental rotation scores and memory capacities, nor in the order in which the treatments are applied, on $i$.\\ \hline

\textbf{H}$\bm{_{i1}}$ &
The treatment (UML-supported requirements vs. textual requirements) has an effect on $i$. \\ \hline

\textbf{H}$\bm{_{i2}}$ &
There is an association between mental rotation ability and $i$. \\ \hline

\textbf{H}$\bm{_{i3}}$ & There is an association between WM capacity and $i$. \\ \hline

\textbf{H}$\bm{_{i4}}$ & The order of treatments (whether UML-supported requirements or textual requirements is applied first) has an effect on $i$. \\ \hline 

\textbf{H}$\bm{_{i5}}$ & The treatment (UML-supported requirements vs. textual requirements) has an effect on $i$ across different levels of mental rotation activity. \\ \hline

\textbf{H}$\bm{_{i6}}$ & The treatment (UML-supported requirements vs. textual requirements) has an effect on $i$ across different levels of WM capacity. \\ \hline

\textbf{H}$\bm{_{i7}}$ & Different levels of WM capacity and mental rotation ability are associated with $i$. \\ \hline

\textbf{H}$\bm{_{i8}}$ & There is a difference in the effect of treatment (UML-supported requirements vs. textual requirements) across different levels of mental rotation scores and WM capacities on $i$.\\ \hline

\end{tabular}

\label{tab:hypotheses}
\end{table*}

\subsection{Analysis Approach}
\subsubsection{Hypotheses}
To answer the RQs, we test a number of null and alternative hypotheses for each dependent variable (cfr. Table \ref{tab:hypotheses}). 

Due to the adopted crossover design, we also test an extra hypothesis related to potential carryover effects (i.e. H$_{i4}$), which does not directly stem from a research question but need to be assessed in this context \cite{vegas2015crossover}.

\subsubsection{Data Analysis}

To test the hypotheses for \textsl{F1-score}, we use a linear mixed-effects model (LMM) 
\cite{vegas2015crossover}. The model includes treatment type (textual vs. UML-supported requirements), cognitive abilities (mental rotation ability and WM capacity), session timing of the inspection task (Day 1 vs. Day 2), and treatment sequence (textual requirements first, followed by UML, or UML first, followed by textual requirements) as fixed effects. Participants were modelled as a random effect to account for individual baseline differences. 
To ensure the appropriateness of the LMM, we verified that the model residuals met the condition of normality by applying the Shapiro-Wilk test \cite{shapiro1965analysis} (W = $0.98692$, p = $0.705$) and visually inspecting residual with normal probability plots. 
The residuals satisfy the normality assumption, supporting the appropriateness of the LMM for our analysis.

To test the hypotheses for \textsl{Accuracy Why}, a generalized linear model 
(GZLM)~\cite{nelder1972generalized} with a quasi-Gaussian family~\cite{wedderburn1974quasi} is used. 
The choice of GZLM was motivated by the nature of the data: \textsl{Accuracy Why} values represent medians from 4-point Likert items, leading to 
non-normally distributed outcomes. Since the assumptions for LMM were not satisfied, we opted for a GZLM using the identity link and quasi-likelihood to handle potential overdispersion and non-normality. 
The model includes as fixed effects treatment type, cognitive abilities, session timing of the inspection task, and treatment sequence, as well as all two- and three-way interaction terms between treatment and cognitive abilities. Unlike the LMMs, this model did not include a random effect, as the GZLM framework does not support random effects in its standard formulation. 
To ensure the appropriateness of the GZLM, we assessed residual distribution through visual inspection and confirmed that the quasi-Gaussian model captured the general trend of the data, despite some deviation from normality. 



To control the risk of Type I error inflation arising from the multiple hypotheses tested within each of our models (i.e., the LMM for \textsl{F1-score}, and the GZLM for \textsl{Accuracy Why}) we adopted a hierarchical, family-based approach to multiplicity adjustment. Although each model included eight hypotheses per dependent variable, not all hypotheses were substantively comparable. Applying a single correction across all eight would have been overly conservative and would have ignored the hierarchical structure of the effects being tested.
To address this, we organised the hypotheses into three families based on the order of interaction:

\begin{itemize}
    \item Family 1: The three-way interaction between treatment, WM capacity, and mental rotation ability (i.e., H$_{i8}$)
    \item Family 2: The two-way interaction (i.e., H$_{i5}$-H$_{i6}$)
    \item Family 3: Main effects of treatment and cognitive abilities (i.e., H$_{i1}$-H$_{i3}$)
\end{itemize}

Within each family, the Benjamini–Hochberg procedure was applied to control the false discovery rate~\cite{benjamini1995controlling}.
This strategy reduces the risk of false positives stemming from testing multiple related hypotheses, while avoiding excessive correction across unrelated effects and maintaining the hierarchical logic of the analysis.

H$_{i4}$ , which tests for carryover effects, serves a different purpose: it assesses the validity of the crossover design rather than the substantive effects of treatment or cognitive abilities. It was, therefore, excluded from the multiplicity adjustment.

To quantify the effect size of each fixed effect in our LMMs, we employed semi-partial R$^2$
~\cite{nakagawa2017coefficient}. 
This approach is particularly well-suited for mixed-effects models, where traditional R$^2$ measures are not directly applicable due to the inclusion of random components.
In our GZLM, we used Nagelkerke’s R$^2$~\cite{nagelkerke1991note}, a pseudo-coefficient of determination specifically designed for models where traditional R$^2$ is not applicable, such as quasi-likelihood models.
Interpretation of these semi-partial R$^2$ values is as follows: 
values less than 2\% are considered negligible, between 2\% and 13\% small, between 13\% and 26\% medium, and greater than 26\% large.

\medskip
In addition to the primary statistical analyses, we conducted a set of post-hoc exploratory analyses to investigate whether participant-related factors might have influenced the outcomes. Specifically, we examined 
(1) whether participants' preferred representation format (UML, text, or both), as indicated in the post-study questionnaire, aligned with the format used during the inspection task; (2) whether prior experience with tasks similar to the mental rotation task was associated with higher rotation scores; and (3) whether participants with a background in technical drawing achieved better results on the rotation task. These exploratory analyses provide additional context to interpret the results. 
The statistical analyses included normality testing (Shapiro-Wilk), non-parametric group comparisons (Wilcoxon rank-sum test), 
and Cohen’s Kappa to evaluate agreement between categorical preferences and observed behaviour. These tests were selected to accommodate the non-normal distribution of several variables and the categorical nature of the data.

\begin{table*}[t]
    \renewcommand{\arraystretch}{1.2}
    \caption{Descriptive statistics.\\{\scriptsize The table reports descriptive statistics for both inspection outcome variables (i.e., \textsl{F1-score}, 
    and \textsl{Accuracy Why}) and covariates (i.e., \textsl{Rotation Score} and \textsl{WM Capacity}). For inspection outcomes variables the table shows the total results (independently from the treatment) in columns denoted with {\bf{\em Total}}; the results for the inspection performed with textual requirements (treatment A) in  columns denoted with {\bf{\em text}}; and the results for the inspection performed with UML-supported requirements (treatment B) in columns denoted with {\bf{\em UML}}.}}
    \label{tab:descrStats}
    \smallskip
    \centering
    \scriptsize
    \begin{tabular}{|c | c |c | c |c | c |c | c | c | c |c | c | c | c |c | c |}
    \hline
      {\bf{\em Variables}}  & \multicolumn{3}{c|}{\bf{\em Median}}  & \multicolumn{3}{c|}{\bf{\em Mean}} & \multicolumn{3}{c|}{\bf{\em Std.\,dev.}} & \multicolumn{3}{c|}{\bf{\em Min.}} & \multicolumn{3}{c|}{\bf{\em Max.}} \\ 
      \hline        \hline
       & \bf{\em Total}  & \bf{\em text} & \bf{\em UML} &  \bf{\em Total}  & \bf{\em text} & \bf{\em UML} & \bf{\em Total}  & \bf{\em text} & \bf{\em UML} & \bf{\em Total}   & \bf{\em text} & \bf{\em UML} & \bf{\em Total}   & \bf{\em text} & \bf{\em UML}  \\ 
      \hline 
      {\bf{\em F1-score}} & 0.476 & 0.487 & 0.467 &  0.464 & 0.471 & 0.457 & 0.130 & 0.145 & 0.116 & 0.175 &   0.175 & 0.197 & 0.733 & 0.733 & 0.655 \\ 
      \hline 
    {\bf{\em Accuracy Why}} & 4 & 4 & 4 & 3.75 & 3.74 & 3.75 & 0.503 & 0.526 & 0.486 & 2 & 2 & 2 & 4 & 4 & 4
 \\ 
      \hline 
      \hline
      
      {\bf{\em Rotation Score}} & \multicolumn{3}{c|}{1} &
      \multicolumn{3}{c|}{0.957} & \multicolumn{3}{c|}{0.089} & \multicolumn{3}{c|}{0.775} & \multicolumn{3}{c|}{1} \\ \hline

      {\bf{\em WM Capacity}} & \multicolumn{3}{c|}{0.763	} &
      \multicolumn{3}{c|}{0.784} & \multicolumn{3}{c|}{0.149} & \multicolumn{3}{c|}{0.451} & \multicolumn{3}{c|}{1} \\ \hline

   \end{tabular}
\end{table*}

\section{Study Results}\label{sect:results}
Table \ref{tab:descrStats} presents the descriptive statistics for both the dependent variable (i.e.,\textsl{F1-score}and \textsl{Accuracy Why}) and the covariates (i.e., \textsl{Rotation Score} and \textsl{WM Capacity}).
Table~\ref{tab:results} summarises the results for all alternative hypotheses. All the p-values reported in the table have been adjusted using the aforementioned Benjamini–Hochberg procedure.

For each dependent variable, we prioritise the interpretation of the three-way interaction term ($H_{i8}$), following the principle that higher-order interactions take precedence in analysis \cite{ellis2010essential}.
Given the exploratory nature of the study and the potential limitations related to sample size, we adopt a significance threshold of $0.1$ for detecting statistical significance.
Furthermore, we consider an effect to be practically significant when it meets two conditions: statistical significance and effect size greater than 2\%.


\subsection{Participants}
In total, 38 participants took part in the study. Given UPM's international environment, all courses are conducted in English and all participants are proficient in English. Additionally,  all participants reported a good level of familiarity with UML sequence diagrams.
%
Participants self-reported their experience with requirements inspection on a 5-point scale: 24\% rated themselves as having little experience (1), 18\% rated 2, 37\% rated 3, and 21\% rated 4; no participants reported level 5.

\subsection{RQ1: Issue detection accuracy}
As Table \ref{tab:descrStats} shows, the mean \textsl{F1-score} across all observations is $0.464$, with a standard deviation of $0.13$, indicating some variability in inspection performance. When broken down by treatment, the mean \textsl{F1-score} for the text group is slightly higher at $0.471$ (SD = $0.145$) compared to the UML group at $0.457$ (SD = $0.116$). This indicates that, on average, \textbf{participants performed slightly better with textual requirements than with UML diagrams}.

Regarding the cognitive variables, the \textsl{Rotation Score} has a mean score of $0.957$ (SD = $0.089$) across all participants. This high mean suggests that participants generally scored well on this metric, showing strong spatial ability. While this implies strong baseline spatial ability in the sample, the lack of variability may limit the ability to discern its influence on inspection accuracy.
\textsl{WM Capacity} has a mean value of $0.784$, with a standard deviation of $0.149$, and a range from $0.451$ to $1$. This indicates variability among participants in WM capacity, which could influence their inspection performance.



As Table~\ref{tab:results} shows, the three-way interaction between treatment, mental rotation ability, and WM capacity (i.e., $H_{FS8}$) is statistically significant (p = 0.0774).
As this is the highest-order interaction considered for this outcome, we prioritise its interpretation over lower-order effects.
The effect size exceeding 2\% indicates as well practical significance.

The negative estimate ($-5.537$) suggests that \textbf{the combined presence of high mental rotation ability and high WM capacity may actually impair performance under the UML treatment.}

\begin{figure}
    \centering
    \includegraphics[width=1\linewidth]{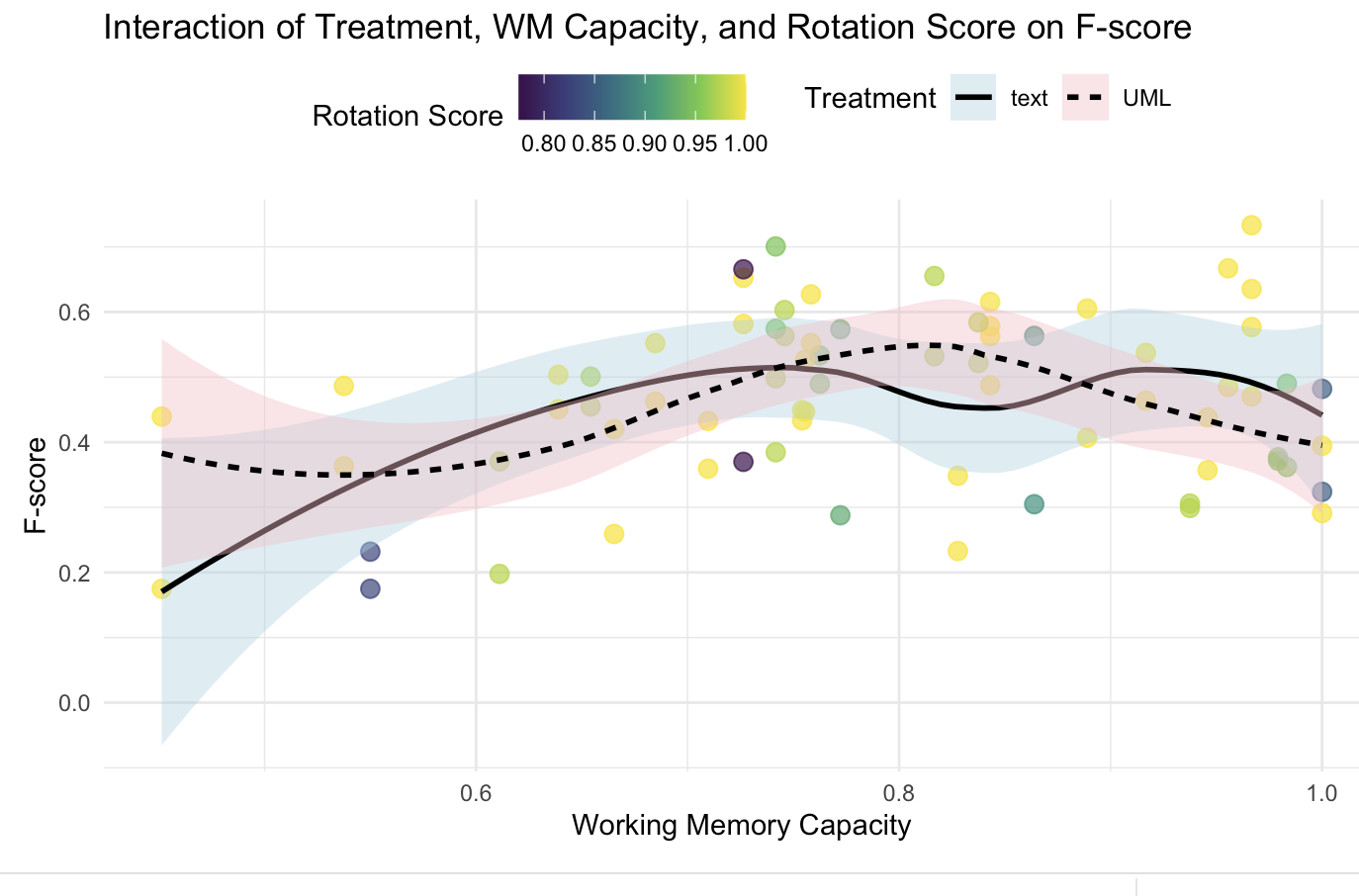}
    \caption{Interaction of Treatment, WM Capacity, and Mental Rotation Ability on \textsl{F1-score}.}
    \label{fig:RQ4f_score}
\end{figure}

As Figure \ref{fig:RQ4f_score} shows, the observed interaction is 
potentially counter-intuitive.
\textsl{F1-score} tends to increase with WM capacity up to a threshold (approximately $0.8$), beyond which performance begins to decline under the UML treatment. This 
trend suggests that while greater WM capacity initially supports better performance under UML-supported requirements, extremely high capacity does not translate into proportionally higher outcomes and may even lead to a subtle performance drop.
The plot also encodes \textsl{rotation score} through the colour gradient of the data points, with higher spatial ability represented by lighter shades (closer to yellow). 
Because nearly all participants have high rotation scores, the colour gradient shows no clear systematic shift along the dashed curve: high-rotation (yellow) and mid-rotation (green) points appear both above and below the\textsl{F1-score}peak (WM Capacity $\approx 0.8$). 
However, it is worth noting that among participants in the declining segment (WM $> 0.8$), those with higher rotation ability (lighter yellow points) are more likely to fall above the trend line, whereas those with lower rotation scores (darker points) tend to fall below. This suggests that 
spatial ability may play a moderating role at the upper end of the WM spectrum---individuals with strong spatial abilities are better able to sustain high accuracy when facing increased information load (UML + text), while those with weaker spatial skills experience a steeper decline.

Overall, the UML curve suggests an ``optimal-capacity window'': \textbf{up to moderate–high WM levels, students leverage their memory resources to offset UML’s cognitive demands, but once that threshold is exceeded, even those with excellent spatial skills experience diminishing returns}.

The random effect for subjects exhibits low variance ($0.0074$) indicating limited variability in baseline performance metrics among participants. This low variance suggests that individual differences in inspection performance are relatively minor and that the majority of variability in these outcomes is explained by the fixed effects, such as treatment type or cognitive abilities, rather than by individual participant differences. 

\vspace{0.2cm}
\hspace*{-5mm}
\begin{tikzpicture}
\node [mybox] (box){%
    \begin{minipage}{.94\columnwidth}
    \centering
\textbf{RQ1. }When WM capacity and mental rotation ability are jointly high, UML support is associated with a decrease in detection accuracy, indicating that UML may introduce cognitive demands that outweigh its benefits beyond an optimal cognitive threshold.
    \end{minipage}
};
\end{tikzpicture}%


\begin{table}[t]
    \renewcommand{\arraystretch}{1.2}
    \caption{Statistical test results addressing the research questions and their corresponding alternative hypotheses. 
    \\{\scriptsize  FS 
    and AW stand for \textsl{F1-score}, 
    and \textsl{Accuracy Why}, respectively. \\ Cells corresponding to practical significant hypotheses (p-value < 0.1 and effect-size > 2\%) are highlighted in grey.\\
    }}
    \label{tab:results}
    \smallskip
    \centering
    \scriptsize
    \vspace*{-0.05cm}
    \begin{tabular}{| p{0.2cm} |c |c | p{0.8cm} | c| p{0.5cm}|}
    \hline 
    & {\bf{\em Hyp.}}   &  {\bf{\em Fixed Effects}}& {\bf{\em Estimate}} & {\bf{\em p-value}} & {\bf{\em Effect size}} \\ \hline \noalign{\vskip\doublerulesep
         \vskip-\arrayrulewidth} 
         \hline

    \multirow{9}{*}[0.2cm]{\rotatebox[origin=c]{90}{RQ1}} & \multicolumn{5}{l|}{\textbf{F1-score}}    \\ \cline{2-6} 
    
    & H$_{\textbf{FS}1}$ & UML & -3.75 & 0.1587 & 2.2\% \\ \cline{2-6}
    &   H$_{\textbf{FS}2}$ & RS & -0.892  & 0.6856 & 0.2\%  \\ \cline{2-6}
    &    H$_{\textbf{FS}3}$ & WM & -0.892  & 0.6856 & 0.4\%  \\ \cline{2-6}
    & H$_{\textbf{FS}4}$ & UML $\rightarrow$ text & 0.0356  & 0.3641 & 1.7\% \\ \cline{2-6}
    & H$_{\textbf{FS}5}$ & UML * RS & 4.017 &  0.1127 & 2.3\% \\ \cline{2-6}
     & H$_{\textbf{FS}6}$ & UML * WM & 5.17 & 0.1127 & 2.4\%  \\ \cline{2-6}
    & H$_{\textbf{FS}7}$ & RS * WM & 1.859 & 0.5181 & 0.6\% \\ \cline{2-6}
     & \cellcolor{gray!15}H$_{\textbf{FS}8}$ & \cellcolor{gray!15}UML * RS * WM & \cellcolor{gray!15}-5.537  & \cellcolor{gray!15}0.0774* & \cellcolor{gray!15}2.6\%  \\ 
     \hline 

    \multirow{9}{*}[0.2cm]{\rotatebox[origin=c]{90}{RQ2}} & \multicolumn{5}{l|}{\textbf{Accuracy Why}}    \\ \cline{2-6} 

   &\cellcolor{gray!15}H$_{\textbf{AW}1}$ & \cellcolor{gray!15}UML & \cellcolor{gray!15}34.13 & \cellcolor{gray!15} 0.0054** &  \cellcolor{gray!15}12.2\% \\ \cline{2-6}
    &     H$_{\textbf{AW}2}$ & RS & 13.69 & 0.1202 &  0.0\%  \\ \cline{2-6} 
    &     H$_{\textbf{AW}3}$ & WM & 18.16 & 0.1202 & 0.0\% \\ \cline{2-6}
     &    H$_{\textbf{AW}4}$ & UML $\rightarrow$ text & 0.05 & 0.656043 &  0.2\% \\ \cline{2-6}

& \cellcolor{gray!15}H$_{\textbf{AW}5}$ & \cellcolor{gray!15}UML * RS & \cellcolor{gray!15}-33.52 & \cellcolor{gray!15}0.0046** &  \cellcolor{gray!15}11\% \\ \cline{2-6}

&    \cellcolor{gray!15} H$_{\textbf{AW}6}$ & \cellcolor{gray!15}UML * WM  & \cellcolor{gray!15}-47.96 & \cellcolor{gray!15}0.0023** & \cellcolor{gray!15}14\% \\ \cline{2-6}
&     H$_{\textbf{AW}7}$ & RS * WM & -16.67 & 0.1127 & 0.0\%  \\ \cline{2-6}
     & \cellcolor{gray!15}H$_{\textbf{AW}8}$ & \cellcolor{gray!15}UML * RS * WM & \cellcolor{gray!15}47.34 & \cellcolor{gray!15}0.0013** &  \cellcolor{gray!15}13\% \\ 
 \hline

   \end{tabular}
\end{table}

\subsection{RQ2: Issue justification accuracy}
\textsl{Accuracy Why} was measured only for TP in issue identification, as it considers only justifications referring to an actual checklist violation; each participant received a single score per condition computed as the median of their justification ratings (on a 1–4 scale). 
The median value for \textsl{Accuracy Why} is $4$ (maximum score of the scale, ``The motivation is fully clear, precise, and strongly aligns with the
error in the requirement, leaving no
ambiguity''), with a standard deviation of approximately $0.5$ (cfr. Table \ref{tab:descrStats}). 
These results are consistent across the different treatments. 
The results suggest that \textbf{once a issue was correctly identified, participants were generally precise and thorough in their justifications, regardless of the treatment condition}. 


As shown in Table~\ref{tab:hypotheses}, the three-way interaction between treatment, WM capacity, and mental rotation ability reveals a significant positive effect of UML diagrams on \textsl{Accuracy Why}  (estimate = $47.34$, $p = 6.50e-04$).
This three-way interaction also represents the second-largest effect size in the model (13\%) indicating as well practical significance.

This suggests that \textbf{the
combined effect of cognitive abilities under UML-supported conditions substantially enhances users’ ability to explain why a requirement violates a checklist item}. Such benefit of UML is not realised through either cognitive ability in isolation, as the two-way interactions between cognitive abilities and UML support (H$_{\textbf{FS}5}$ and H$_{\textbf{FS}6}$) show a significant negative effect.

This pattern suggests a cognitive fit effect: \textbf{UML enhances reasoning quality only when users possess both sufficient WM capacity and mental rotation skills, while in cases of imbalance, the added cognitive load of reading diagrams may outweigh the benefits}.

\begin{figure}
    \centering
    \includegraphics[width=1\linewidth]{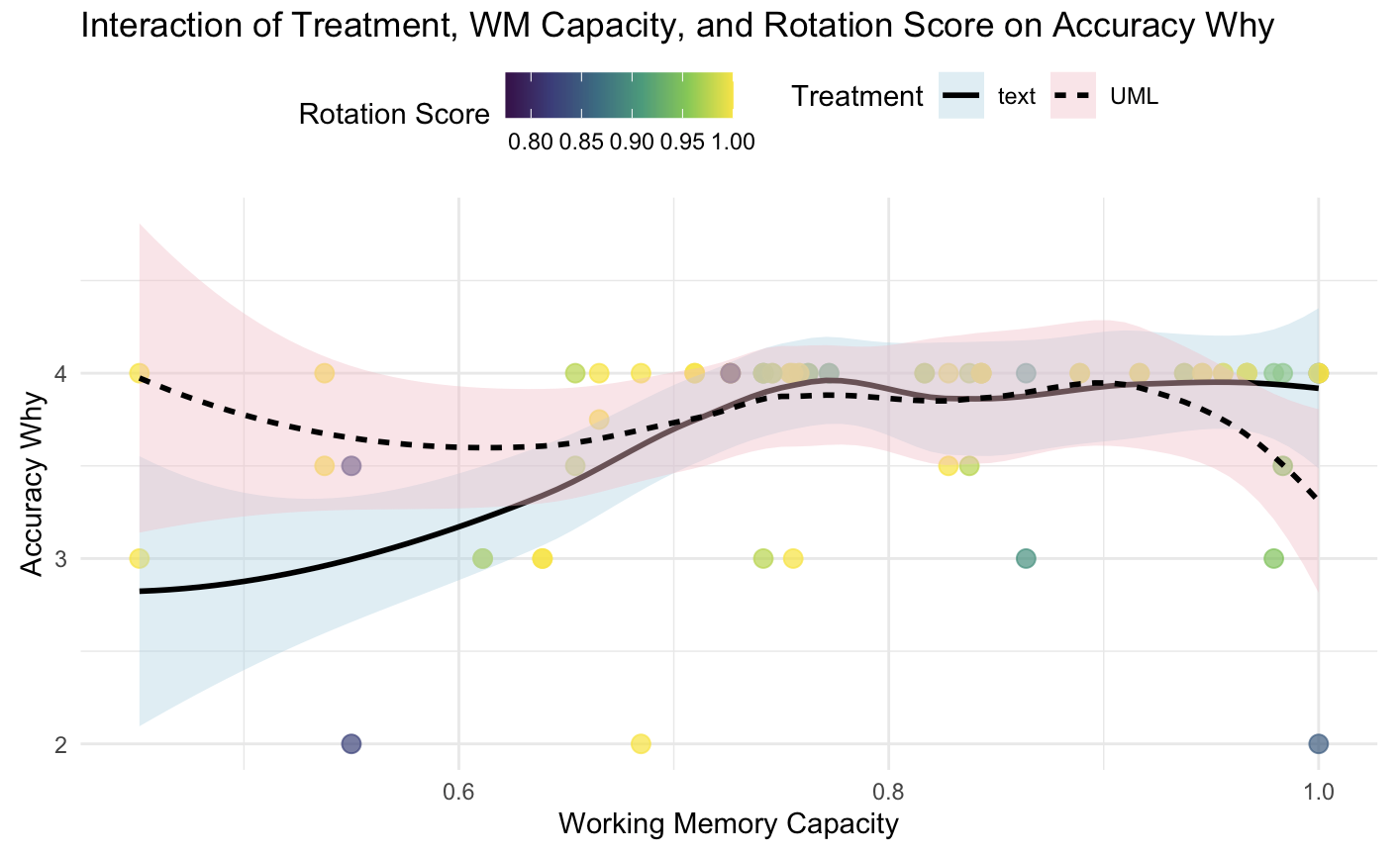}
    \caption{Interaction of Treatment, WM Capacity, and Mental Rotation Ability on \textsl{Accuracy Why}.}
    \label{fig:RQ4accuracyWhy}
\end{figure}

As in the case of \textsl{F1-score}, the interaction between treatment, WM capacity, and mental rotation ability exhibits a counter-intuitive  pattern for \textsl{Accuracy Why} (cfr. Figure \ref{fig:RQ4accuracyWhy}).
Under the text treatment (solid black line), justification accuracy
increases with WM capacity up to approximately $0.8$, beyond which it
remains relatively stable. In contrast, under the UML treatment (dashed black line), accuracy remains approximately stable until WM capacity reaches a value of $\approx0.9$, beyond which it declines.

It is worth noting that mental rotation scores exhibit low variability, resulting in a concentration of data points toward the lighter end of the colour gradient (closer to yellow), indicating that most participants had high spatial ability. However, below the declining segment of the UML curve (WM $> 0.9$) data points are darker, suggesting that at very high levels of WM capacity, lower spatial abilities may hinder participants’ ability to provide accurate justifications. 
This pattern aligns with the significant three-way interaction observed in the model, where the benefit of UML emerges only when both cognitive abilities are present in balance. The trend further underscores that cognitive support \textbf{tools like UML are not universally beneficial---they may enhance reasoning for some users but become counterproductive for others, depending on their individual cognitive profile}.

\vspace{0.2cm}
\hspace*{-5mm}
\begin{tikzpicture}
\node [mybox] (box){%
    \begin{minipage}{.95\columnwidth}
    \centering
\textbf{RQ2.} UML enhances the quality of justifications under balanced cognitive abilities, whereas no advantage---and even negative effects---are observed when cognitive abilities are considered in isolation or are unbalanced.
    \end{minipage}
};
\end{tikzpicture}%

\subsection{Carryover Effect}
The results on H$_{i4}$ for both \textsl{F1-score} and \textsl{Accuracy Why} (cfr. Table \ref{tab:results}) revealed no significant carryover effects, with p-values well above typical significance thresholds. This suggests that \textbf{participants’ performance did not change depending on the sequence in which they encountered the UML and text treatments}.

\section{Post-hoc Analysis}\label{sect:postHoc}
To explore the potential influence of participant-related factors on inspection outcomes and cognitive measures, we conducted a series of post-hoc analyses. Specifically, we examined whether the high rotation scores observed could be attributed to prior exposure to similar tasks or experience with technical drawing. 
We also assessed whether participants’ stated preferences for representation format aligned with their actual behaviour during the inspection tasks. 

\textbf{Used vs. preferred representation format.}
To assess whether participants used the representation format they ultimately preferred, we compared their stated preference (reported after the study as part of the post-test questionnaire) with the format most frequently used during the inspection task. Only 14\% used the format matching their stated preference, while 86\% showed a mismatch. Cohen’s Kappa was computed to measure agreement between preference and usage, as performed in \cite{vegas2020mis}. The result was $K = 0.0508$, with $p = 0.137$, indicating very weak agreement, yet no significant. This suggests that \textbf{participants’ retrospective preferences may not accurately reflect their behaviour during the task}. 

\textbf{\textsl{Rotation score} and prior task experience.}
A Wilcoxon rank-sum test was conducted to determine whether rotation scores differed between participants who had previously performed similar tasks and those who had not. The test revealed no statistically significant difference in rotation scores between the two groups ($W = 157.5$, $p = 0.4787$). This suggests that \textbf{prior experience with similar tasks does not explain the high and homogeneous rotation scores observed in the sample}.

\textbf{\textsl{Rotation score} and technical drawing background.}
We applied a Wilcoxon rank-sum test to assess whether participants with a background in technical drawing exhibited higher rotation scores. The results indicated no significant difference in rotation scores between those with and without such experience ($W = 198.5$, $p = 0.4737$). As with the previous analysis, this suggests that \textbf{technical drawing expertise does not account for the limited variability in spatial ability observed}.



\section{Discussion}\label{sect:discussion}
The results obtained indicate that, within the context of this study, the effect of UML diagrams on inspection performance depends on how cognitive abilities are considered.
When the interaction between UML and each cognitive ability is considered separately (H${i5}$, H${i6}$),
higher WM capacity and mental rotation ability are associated with a slight (though non-significant) improvement in identification accuracy (\textsl{F1-score}), but a significant reduction in justification accuracy (\textsl{Accuracy Why}). 
On the contrary, when UML use and both cognitive abilities are analysed jointly (H$_{i8}$), UML diagrams decrease the ability to identify violations but increase the ability to justify them. 
This suggests that the combination of high spatial and WM abilities may lead to cognitive overload during the detection task, yet support deeper reasoning when justifying violations. UML diagrams may require students to hold and manipulate complex representations, increasing cognitive demands and possibly pushing high-capacity individuals beyond optimal load levels during inspection tasks.
Cognitive abilities appear to play a dual role. On the one hand, they serve as compensatory resources during more complex reasoning tasks---such as articulating justifications---where they support clearer argumentation. On the other hand, they may backfire during violations identification under cognitively demanding conditions like UML interpretation.  

Looking more in detail at the 3-way plots for both \textsl{F1-score} and \textsl{Accuracy Why} (cfr. Figures \ref{fig:RQ4f_score} and \ref{fig:RQ4accuracyWhy}), a consistent pattern emerges: after a certain threshold in WM capacity---approximately >0.8 for \textsl{F1-score} and >0.9 for \textsl{Accuracy Why}---performance under the UML condition begins to decline.
Mental rotation ability, while uniformly high in this cohort, does not substantially alter that turning point: it cannot fully rescue performance when WM saturation occurs with UML. 

The limited variability in mental rotation scores represents a critical limitation of the study. 
The concentration of scores at the upper end of the scale reduces the ability to detect potential effects of mental rotation ability on inspection outcomes. Consequently, both the direct influence of \textsl{rotation score} and its interactions with other factors---such as treatment and \textsl{WM capacity}---are difficult to interpret reliably. This constraint suggests the need for future studies to include a more diverse sample in terms of spatial ability.

Nonetheless, these observations highlight a complex trend in how cognitive abilities relate to inspection performance, cautioning 
against assuming that ``more ability always means better performance''. 
For student populations in particular, it is crucial to balance task complexity with cognitive load---especially when introducing visual notations such as UML that place additional demands on WM.

One plausible explanation of 
this trend
lies in the concept of cognitive overload. Participants with high WM capacity may be more adept at managing the visual and structural complexity of UML diagrams, but those with exceptionally high capacity might over-engage with the material. They may attempt to retain and process excessive details simultaneously, inadvertently increasing their cognitive load beyond optimal levels. According to Cognitive Load Theory \cite{sweller2011cognitive}, performance is maximised at an optimal load level; when this is exceeded, additional cognitive resources may yield diminishing returns or even impair performance.

This phenomenon is further supported by psychological research. For instance, individuals with higher WM capacity may exhibit overconfidence in their ability to handle complex information, leading them to persist in using cognitively demanding strategies that are not always effective for the task at hand \cite{decaro2016higher}. Furthermore, Dual-Process Theory posits that individuals with higher WM capacity are more inclined to engage in controlled (Type 2) processing, which is slower and effortful \cite{evans2013dual}. Under high cognitive load, such as interpreting UML diagrams alongside textual specifications, this type of reasoning becomes less efficient and may hinder overall performance \cite{lavie2010attention}.

Another contributing factor could be resource allocation and attentional control. High-capacity individuals might attempt to process all available information---both textual and visual---leading to fragmented attention and greater susceptibility to cognitive interference. Research suggests that these individuals may, paradoxically, be more distractible under high WM load, especially in complex multitasking contexts \cite{ahmed2012focusing}.

These results highlight not only the importance of tailoring notations and tools to users’ cognitive profiles, but also the need to consider cognitive abilities as a whole rather than in isolation. 
Particularly in educational or novice settings, UML diagrams may demand scaffolding strategies---such as progressive disclosure, integrated guidance, or cognitive training---to manage the cognitive demands of UML and avoid overloading even those with strong cognitive abilities. 

Our findings partly align with those of Sharif et al.~\cite{sharifexamining}, who assessed WM capacity and mental rotation independently and found only weak and inconsistent associations with defect detection accuracy across different UML layouts, consistent with our own observation that analysing abilities in isolation offers limited explanatory power.
By modelling the interaction between cognitive abilities, our study reveals effects that remain hidden when each ability is analysed independently, underscoring the importance of considering cognitive abilities as an integrated system when assessing inspection performance.

\section{Threats to Validity}\label{sect:threats}
This section discusses the potential threats to the validity of our study organised according to \cite{wohlin2012experimentation}.

\subsection{Construct Validity}
 In software inspection studies, the success of the inspection phase is typically quantified by the number of violations identified \cite{mendoncca2008framework,rong2012effect} or by calculating the average number of violations detected against the total number of violations, alongside the associated costs for their identification \cite{laitenberger2000experimental,sabaliauskaite2002experimental}. In our study, we measured accuracy using \textsl{F1-score} to account for both false positives and false negatives, providing a more nuanced evaluation of inspection performance. While these metrics are standard in information retrieval and violation detection, they may not fully capture the nuances of requirements inspection, such as the severity or complexity of identified violation.
 To address this limitation, we introduced \textsl{Accuracy Why}, a complementary measure that evaluates the quality of participants’ justifications for identified violations. Justifications were scored on a 0–4 Likert scale by two of the authors based on a predefined rubric assessing completeness, correctness, and clarity of reasoning
 , with discrepancies resolved through discussion to ensure consistency. This approach extends traditional inspection metrics by capturing not only whether a violation was found, but also how well participants could articulate why it constituted a violation.

For cognitive ability assessments, we employed established measures: mental rotation ability was evaluated using a 3D mental rotation task, while WM capacity was measured via an operation span task. Although these instruments are validated in cognitive psychology and software engineering research contexts (see e.g. \cite{mansoor2023empirical}), the high homogeneity in rotation scores among our participants (mean = 0.957, SD = 0.089) limited our ability to explore the full spectrum of spatial ability effects.

The treatment representations---UML diagrams and textual requirements---were designed to be equivalent in content, but variations in participants’ familiarity with UML or perceived complexity could have introduced bias. To mitigate this, future studies could include pre-study assessments of UML proficiency and use more diverse task materials to better represent real-world requirements documents.

The checklist used to guide violation identification is a simplified version of the checklist used by Porter et al.~\cite{porter1995comparing}; while this simplification may reduce coverage and nuance compared to the original, it improves usability for student participants.

\subsection{Internal Validity}
Learning effects are a potential concern, as participants completed inspections in two sessions. Although a one-week washout period was applied, some residual learning could still have occurred. To check this, we tested the carryover hypothesis (H$_{i4}$), which showed no statistically significant effect, indicating that carryover did not meaningfully influence the results.

Another threat concerns the timing of the cognitive tests, which participants completed online at their convenience. This may have introduced variability due to differences in fatigue or environmental distractions. Although groups were balanced on cognitive abilities, unmeasured factors such as motivation or attention could still have influenced performance. Future replications could standardise testing conditions and include additional checks on participant engagement.

\subsection{External Validity}
The study involved master’s students in computer engineering, many of whom have some industrial experience; although students are recognised to be adequate proxy for professionals in SE experiments \cite{falessi2018empirical}
, their level of professional practice may still differ from that of full-time industry professionals. Students’ experience with inspections or UML could skew results compared to experienced practitioners. Additionally, the inspection tasks focused on video game requirements, which may not reflect the complexity or domain-specific challenges of real-world projects.
To enhance external validity, future work could replicate the study with professionals from varied industries and incorporate more diverse requirement types.

\subsection{Conclusion Validity}
Although the sample size of 38 participants may have reduced the power to detect small effects
the study was still able to identify meaningful and statistically significant interaction patterns.
The \textsl{rotation score}, measured after participant recruitment, exhibited high homogeneity, making it difficult to assess its full impact. Pre-screening for cognitive variability could address this in future studies.

\section{Conclusions}
\label{sect:conclusion}
This study investigated the impact of cognitive abilities---WM capacity and mental rotation ability---on the accuracy of requirements inspection tasks performed using UML versus textual representations. Through a controlled crossover design involving student participants, we examined both identification and justification accuracy.

Our findings highlight a complex interaction between representation type and cognitive abilities. While higher cognitive abilities in isolation were associated with improved issue detection under UML support, a significant three-way interaction revealed that their combined effect could hinder performance in identification tasks, likely due to cognitive overload. Conversely, the same combination facilitated better justifications, suggesting a compensatory role of cognitive abilities in more reflective reasoning tasks. These results caution against assuming that stronger cognitive abilities always translate into better performance, particularly in cognitively demanding activities like UML-based inspections.

The findings also underline the importance of tailoring representations and tools to users' cognitive profiles. UML diagrams, while potentially beneficial for justification, may introduce an additional cognitive burden during issue detection, especially among users with strong but possibly misaligned strategies. In educational or novice settings, this may call for additional support mechanisms to scaffold UML interpretation and prevent overload.

Several directions can extend this line of research. The limited variability in mental rotation scores among participants constrained the generalisability of our results. Future studies should involve more heterogeneous samples, including professionals with varied experience levels and cognitive profiles. 
Future work could also explore the development of adaptive or personalised inspection environments that adjust the level of visual complexity based on real-time assessments of cognitive load or user profiles. 
Finally, extending this research to industrial settings, where the level of domain expertise differ significantly, would help validate the ecological validity of the observed effects.






\bibliographystyle{IEEEtran}
\bibliography{biblio}

\end{document}